\pdfoutput=1  
\documentclass[]{article}  
\usepackage{url,float}
\usepackage{graphicx}
\usepackage{amsfonts}
\usepackage{amssymb}
\usepackage{latexsym}

\newcommand{\hide}[1]{}

\newcommand{\ABox}{
\raisebox{3pt}{\framebox[6pt]{\rule{6pt}{0pt}}}
}
\newenvironment{proof}{{\bf Proof:}}{\hfill\ABox}

\newtheorem{theorem}{{\bf Theorem}}

\newtheorem{lemma}[theorem]{Lemma}

\newcommand{\lemlab}[1]{\label{lemma:#1}}
\newcommand{\thmlab}[1]{\label{thm:#1}}

\newcommand{\figlab}[1]{\label{fig:#1}}
\newcommand{\seclab}[1]{\label{sec:#1}}

\newcommand{\lemref}[1]{\ref{lemma:#1}}
\newcommand{\thmref}[1]{\ref{thm:#1}}

\newcommand{\eqref}[1]{\ref{eq:#1}}
\newcommand{\figref}[1]{\ref{fig:#1}}

{\makeatletter
 \gdef\xxxmark{%
   \expandafter\ifx\csname @mpargs\endcsname\relax 
     \expandafter\ifx\csname @captype\endcsname\relax 
       \marginpar{xxx}
     \else
       xxx 
     \fi
   \else
     xxx 
   \fi}
 \gdef\xxx{\@ifnextchar[\xxx@lab\xxx@nolab}
 \long\gdef\xxx@lab[#1]#2{{\bf [\xxxmark #2 ---{\sc #1}]}}
 \long\gdef\xxx@nolab#1{{\bf [\xxxmark #1]}}
 \gdef\turnoffxxx{\long\gdef\xxx@lab[##1]##2{}\long\gdef\xxx@nolab##1{}}%
}

\def\P{{\mathcal P}}

\def\G{{\Gamma}}

\def\o{{\omega}}

\def\s{{\sigma}}

\def\a{{\alpha}}
\def\b{{\beta}}
\def\sp{\mathop{\rm sp}\nolimits}
\def\bP{{\partial P}}

\newcommand{\squeezelist}{\setlength{\itemsep}{0pt}}

\title{
Unfolding Convex Polyhedra \\ 
via
Quasigeodesic Star Unfoldings\footnote{
   A preliminary version of this work appeared
   in~\cite{iov-ucpq-07,iov-ucpq-07a}.
}
}

\author{%
Jin-ichi Itoh%
    \thanks{Dept. Math.,
	Faculty Educ., Kumamoto Univ.,
	Kumamoto 860-8555, Japan.
    \protect\url{j-itoh@kumamoto-u.ac.jp}}
\and
Joseph O'Rourke%
    \thanks{Dept. Comput. Sci., Smith College, Northampton, MA
      01063, USA.
      \protect\url{orourke@cs.smith.edu}.}
\and
Costin V\^{i}lcu%
    \thanks{Inst. Math. `Simion Stoilow' Romanian Acad.,
P.O. Box 1-764,
RO-014700 Bucharest, Romania.
    \protect\url{Costin.Vilcu@imar.ro}.}
}

\begin{document}
\maketitle

\begin{abstract}
We extend the notion of a star unfolding to be based on
a simple quasigeodesic loop $Q$ rather than on a point.
This gives a new general method to unfold the surface of any convex polyhedron
$\P$ to a simple, planar polygon: shortest paths
from all vertices of $\P$ to $Q$ are cut, and all but one
segment of $Q$ is cut.
\end{abstract}

\section{Introduction}

There are two general methods known to unfold the surface $\P$ 
of any convex polyhedron
to a simple polygon in the plane:
the source unfolding and the star unfolding.
Both unfoldings are with respect to a point $x \in \P$.
Here we define a third general method:
the star unfolding
with respect to a
simple closed ``quasigeodesic loop'' $Q$ on $\P$.
In a companion paper~\cite{iov-ucpqsu-08b}, we 
will extend the analysis
to the source unfolding with respect to a wider class of curves $Q$.

The \emph{point source unfolding} cuts the \emph{cut locus} of
the point $x$:
the closure of set of all those points $y$ to which there is
more than one shortest path on $\P$ from $x$.
Alternatively, the cut locus is the set of all extremities
(different from $x$) of maximal (with respect to inclusion) shortest paths
starting at $x$.
The notion of cut locus was introduced by 
Poincar\'e~\cite{p-lgsc-1905}
in 1905, and since then has gained an
important place in global Riemannian geometry; see, e.g., 
\cite{k-ccl-67} or~\cite{s-rg-96}.
The point source unfolding has been studied
for polyhedral surfaces since~\cite{ss-spps-86} 
(where the cut locus is called the ``ridge tree'').
The \emph{point star unfolding} cuts the shortest paths from $x$ to every
vertex of $\P$.
The idea goes back to Alexandrov~\cite[p.~181]{a-kp-48};%
\footnote{
   It is called the ``Alexandrov unfolding'' in~\cite{mp-mccpc-05}.
}
that it unfolds $\P$ to a simple polygon was established
in~\cite{ao-nsu-92}.

In this paper we extend the star unfolding to be based on a
simple closed polygonal curve $Q$ with particular properties,
rather than based a single point.
This unfolds any convex polyhedron to a simple polygon,
answering a question
raised in~\cite[p.~307]{do-gfalop-07}.

The curves $Q$ for which our star unfolding works are
quasigeodesics, which we now define.


\paragraph{Geodesics \& Quasigeodesics.}
A \emph{geodesic} is a locally shortest path on a smooth surface.
A quasigeodesic is a generalization that extends the notion to 
nondifferentiable, and in particular, to
polyhedral surfaces.
Let $\G$ be any directed curve on a convex surface $\P$, 
and $p \in \G$ be any point in the relative interior of $\G$, i.e., not
an endpoint.
Let $L(p)$ be the total face angle incident to the left side of $p$,
and $R(p)$ the angle to the right side.
If $\G$ is a geodesic, then $L(p){=}R(p)=\pi$.
A \emph{quasigeodesic} $\G$ loosens this condition to $L(p) \le \pi$ and $R(p) \le \pi$,
again for all $p$ interior to $\G$~\cite[p.~16]{az-igs-67}~\cite[p.~28]{p-egcs-73}.
So a quasigeodesic $\G$ has $\pi$ total face angle incident to each
side at all nonvertex points (just like a geodesic),
and has ${\le}\pi$ angle to each side where $\G$ passes through a
polyhedron vertex.
(Geodesics can never pass through vertices.)
A \emph{simple closed geodesic} is non-self-intersecting (\emph{simple}) closed
curve that is a geodesic, and
a \emph{simple closed quasigeodesic} is a simple closed curve
on $\P$ that is quasigeodesic throughout its length.
As all curves we consider must be simple, we will henceforth
drop that prefix.
Pogorelov showed that any convex polyhedron $\P$ has at least three closed
quasigeodesics~\cite{p-qglcs-49}, extending the celebrated earlier result of
Lyusternik-Schnirelmann showing that the same holds for geodesics on 
differentiable convex surfaces.

A \emph{geodesic loop} is a closed curve that is geodesic
everywhere except possibly at one point,
and similarly a \emph{quasigeodesic loop} is quasigeodesic except
possibly at one point $x$, the \emph{loop point}, at which the angle conditions
on $L(x)$ and $R(x)$ may be violated---one may be ${>} \pi$.
Quasigeodesic loops encompass
closed geodesics and quasigeodesics,
as well as geodesic loops.

Although it is known that every $\P$ must have at least
three closed quasigeodesics, 
there is no algorithm known that will find a simple closed quasigeodesic
in polynomial time:
Open Problem~24.2~\cite[p.~374]{do-gfalop-07}.
Fortunately it is in general
easy to find quasigeodesic loops on a given $\P$:
start at any nonvertex point $p$, and extend a geodesic from
$p$ in opposite directions, following each branch until they meet
at $x$.  If no vertices are encountered, we have a geodesic loop;
if vertices are encountered, maintaining the angle conditions through the
vertices (which is always possible, e.g., by bisecting the surface angle) will
result in a quasigeodesic loop.

An exception to this ease of finding a geodesic loop could
occur on an \emph{isosceles tetrahedron}: a tetrahedron
whose four faces are congruent triangles, or, equivalently,
one at which the total face angle incident to each vertex is $\pi$.
It is proved in~\cite{iv-gcit-08} that a convex surface possesses
a \emph{simple quasigeodesic line}---a
non-self-intersecting quasigeodesic infinite in both directions---if
and only if the surface is an isosceles tetrahedron.
So, excepting this case, the procedure described above will produce a
quasigeodesic loop.

\paragraph{Discrete Curvature.}
The discrete \emph{curvature} $\o(p)$ at any point $p \in \P$,
is the \emph{angle deficit}
or \emph{gap}: $2\pi$ minus the sum of the face angles incident to $p$.
The curvature is only nonzero at vertices of $\P$;
at each vertex it is positive because $\P$ is convex.
By the Gauss-Bonnet theorem, a closed geodesic partitions the
curvature into $2\pi$ in each ``hemisphere'' of $P$.
For quasigeodesics that pass through vertices, the curvature in each
half is ${\le}2\pi$.
The curvature in each half defined by a quasigeodesic loop depends on the
angle at the exceptional loop point.

\paragraph{Some Notation.}
For a quasigeodesic loop $Q$ on $\P$,
$\P \setminus Q$ separates $\P$ into two ``halves''
$P_1$ and $P_2$.
As our main focus is usually on one such half, 
to ease notation we sometimes use $P$ without a subscript
to represent either of $P_1$ or $P_2$
when the distinction between them does not matter.
Unless otherwise stated,
vertices of $\P$ are labeled $v_i$ in arbitrary order.
Other notation will be introduced as needed.
A glossary of all symbols defined (chronologically) 
throughout the paper is provided in Appendix~1.


\section{Example of Star Unfolding}
\seclab{Example}
We start with an example.
Figure~\figref{GeodesicLoopCube}(a) shows a geodesic loop
$Q$ on the surface $\P$ of a cube.
$L(p){=}R(p)=\pi$ at every point $p$ of $Q$ except at
$x$, where $R(x){=}\frac{3}{2}\pi$ and  $L(x){=}\frac{1}{2}\pi$.
Note that three cube vertices, $\{v_3,v_6,v_7\}$, are to the left
of $Q$, and the other five to the right.
This is consistent with the Gauss-Bonnet theorem, because
$Q$ has a total turn of $\frac{1}{2}\pi$,
so turn plus enclosed curvature is $2\pi$.

For each vertex $v_i \in \P$, we select a shortest path
$\sp(v_i)$ to $Q$: a geodesic from $v_i$ to a point $v'_i \in Q$
whose length is minimal among all geodesics to $Q$.
In general there could be several shortest paths from $v_i$ to $Q$;
we use $\sp(v_i)$ to represent an arbitrarily selected one.
The point $v'_i \in Q$ is called
a \emph{projection} of $v_i$ onto $Q$,
and the path $\sp(v_i) = v_i v'_i$ is called a \emph{segment} on $\P$.
In the example, all the shortest path segments $\sp(v'_i)$ are unique,
which is the generic situation.

\begin{figure}[htbp]
\centering
\includegraphics[width=\linewidth]{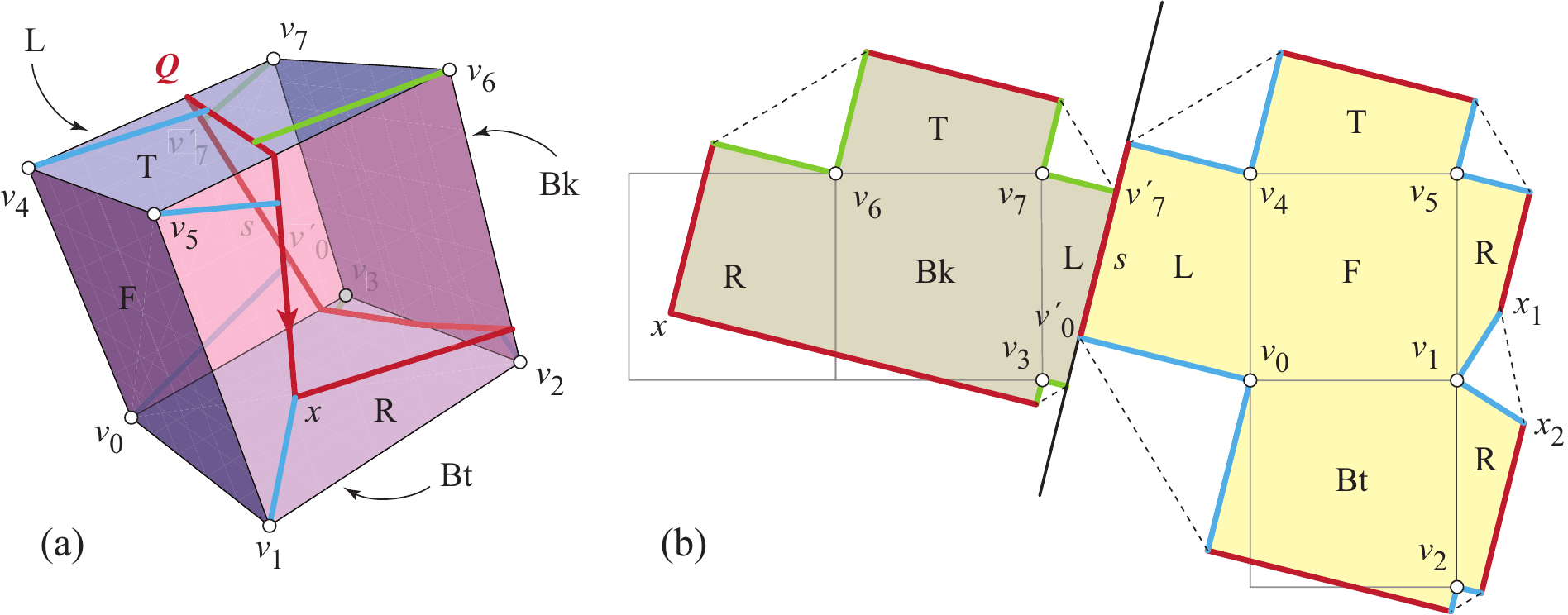}
\caption{
(a)~Geodesic loop $Q$ on cube.
Shortest paths $\sp(v_i)$ are shown.  Faces are labeled
$\{F,T,L,R,Bt,Bk\}$.
(b)~Star unfolding with respect to $Q$,
joined at 
$s=v'_0 v'_7$.
}
\figlab{GeodesicLoopCube}
\end{figure}

\paragraph{Algorithm.}
If we view the star unfolding as an algorithm
with input $\P$ and $Q$, it consists of three
main steps:
\begin{enumerate}
\squeezelist
\item Select shortest paths $\sp(v_i)$ from each $v_i \in \P$ to $Q$.
\item Cut along $\sp(v_i)$ and flatten each half.
\item Cut along $Q$, joining the two halves at an uncut segment $s \subset Q$.
\end{enumerate}

After cutting along $\sp(v_i)$, we conceptually insert an isoceles
triangle with apex angle $\o(v_i)$ at each $v_i$, which flattens
each half.  One half (in our example, the left half), is convex,
while the other half has several points of nonconvexity,
at the images 
of $x$.
(In our example, only $x_1$ is nonconvex, when the inserted
``curvature triangles'' are included.)
In the third and final step of the procedure,
we select a segment $s$ of $Q$ whose interior contains
neither a vertex $v_i$ nor any vertex projection $v'_i$, 
such that the extension of $s$ is a supporting line of each half,
and cut all of $Q$ except for $s$.
In our example, we choose 
$s=v'_0 v'_7$
(many choices for $s$ work in this example),
which leads to non-overlap of the two halves.

We now proceed to detail the three steps of the procedure,
this time with proofs.  We use a different example
to illustrate the main ideas.

\section{Shortest Path Cuts}
We again use
a cube as an illustrative example, but this time with a 
closed quasigeodesic $Q$, not a loop: 
$Q=(v_0,v_5,v_7)$; see
Figure~\figref{cube_geodesic_hemis}(a).
There is $\pi$ angle incident to the right at $v_5$, 
and $\frac{1}{2}\pi$ incident to
the left; and similarly at $v_0$ and $v_7$.  At all other
points $p \in Q$, $L(p){=}R(p)=\pi$.  Thus $Q$ is indeed a quasigeodesic.
We will call the left half (including $v_2$) $P_1$, and the right half 
(including $v_4$) $P_2$.
In Figure~\figref{cube_geodesic_hemis}(a), the paths from $\{v_1,v_3,v_6\}$
are uniquely shortest.
From $v_2$ there are three paths tied for shortest, and from
$v_4$ also three are tied.
\begin{figure}[htbp]
\centering
\includegraphics[height=0.95\textheight]{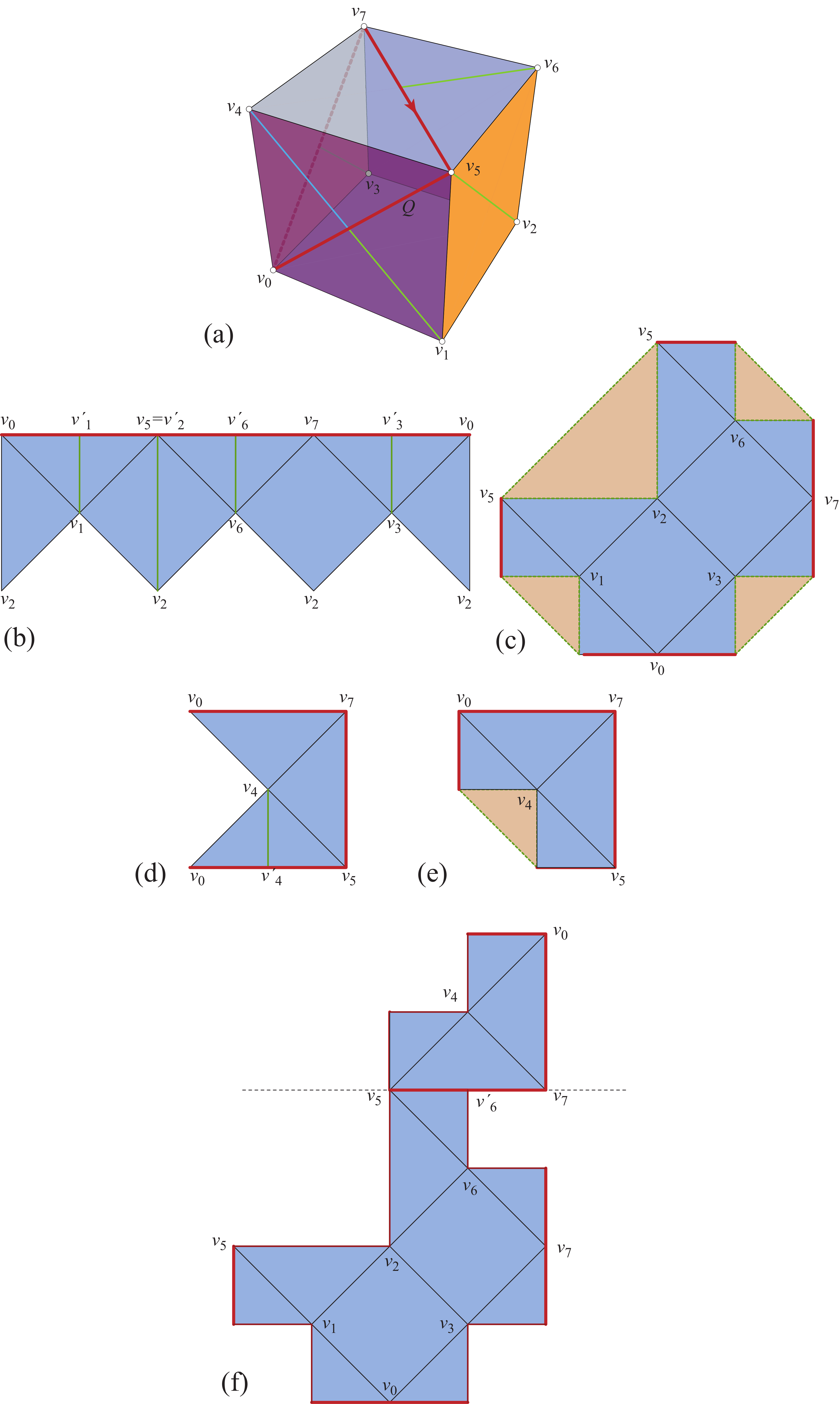}
\caption{
(a)~Cube and quasigeodesic $Q=(v_0,v_5,v_7)$. 
Shortest paths $\sp(v_i)$ as indicated.
(b,c) Flattening the left half  by insertion of curvature triangles
along the shortest paths $\sp(v_i)=v_iv'_i$.
(d,e) Flattening the right half.
(f)~Two halves joined at $s=v_5 v'_6$.
}
\figlab{cube_geodesic_hemis}
\end{figure}

A central fact that enables our construction is this key lemma from
\cite[Cor.~1]{iiv-qfpcs-07}, slightly modified for our circumstances:%

\begin{lemma}                                                                   
Let $Q$ be a quasigeodesic on a convex surface $\P$, and $p$ any point of       
$\P$ not on $Q$.                                                                
Then for any choice of $\sp(p)=pp'$, this is the unique shortest path from      
$p$ to $p'$ and it is orthogonal to $Q$ if $p'$ is in the relative              
interior of $Q$.                                                                
\lemlab{IIV}                                                                    
\end{lemma}

In our situation, the orthogonality condition is only guaranteed
to hold when $p'$ is not the exceptional loop point $x$ of a quasigeodesic
loop.  In the example of Figure~\figref{cube_geodesic_hemis}(a),
there is no exceptional point, so all projections are orthogonal
to $Q$.
Note that this lemma does
not say that the shortest path from $p$ to $Q$ is unique---which we
know is not always true---but that,
among those that are tied for shortest, each is the unique shortest path
between its two endpoints.

A second fact we need concerning these shortest paths is that they are disjoint,
excepting those arriving at the exceptional loop point,
in which case they share precisely that point.
The reader who accepts this basic fact is invited to skip beyond
the proof.
\begin{lemma}
Any two shortest paths 
$\sp(v_1)$ and $\sp(v_2)$,
not incident to the loop point $x$,
are disjoint,
for distinct vertices $v_1,v_2 \in P$.
\lemlab{sp.disjoint}
\end{lemma}
\begin{proof}
Suppose for contradiction that at least one point $u$ is shared:
$u \in \sp(v_1)\cap \sp(v_2)$.
We consider four cases:
one shortest path is a subset of the other,
the shortest paths cross, the shortest paths touch at an interior point
but do not cross, or their endpoints coincide.
\begin{enumerate}
\item $\sp(v_2) \subset \sp(v_1)$.
Then $\sp(v_1)$ contains a vertex $v_2$ in its interior, which violates
a property of shortest paths~\cite[Lem.~4.1]{ss-spps-86}.
\item $\sp(v_1)$ and $\sp(v_2)$ cross properly at $u$.
It must be that $|uv'_1|=|uv'_2|$, otherwise both paths
would follow whichever tail is shorter.
But now it is possible to shortcut the path in the vicinity of $u$ via
$\s$ as shown in 
Figure~\figref{sp_lemma}(a), and the path $(v_1,\s,v'_2)$ is shorter than $\sp(v_1)$.
\item $\sp(v_1)$ and $\sp(v_2)$ touch at $u$ but do not cross properly there.
Then there is a shortcut $\s$ to one side (the side with angle ${<}\pi$),
as shown in 
Figure~\figref{sp_lemma}(b).
\item $v'_1 = v'_2$.
Then from Lemma~\lemref{IIV}, we know the two paths are orthogonal to the
quasigeodesic $Q$.  If we are not in the previous case, then it
must be that there is an angle $\a > 0$ separating the paths in a neighborhood
of the common endpoint; see Figure~\figref{sp_lemma}(c).
Then $Q$ has more than $\pi$ angle to one side at
this point, violating the definition of a quasigeodesic.
Note that it is here we use the assumption that the paths
are not incident to the loop point $x$.
\end{enumerate}
\end{proof}

\begin{figure}[htbp]
\centering
\includegraphics[width=\linewidth]{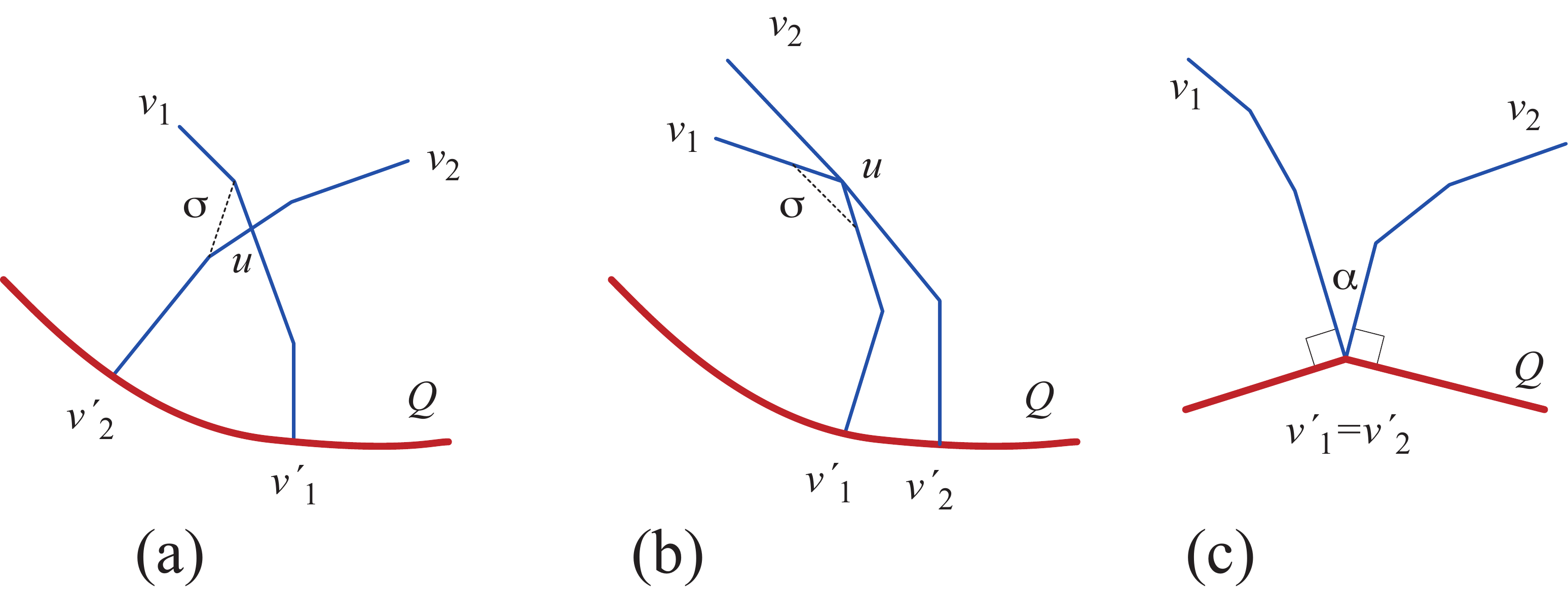}
\caption{Lemma~\protect\lemref{sp.disjoint}:
(a)~paths cross;
(b)~paths touch at an interior point;
(c)~paths meet at endpoint.}
\figlab{sp_lemma}
\end{figure}

This lemma ensures that the cuttings along $\sp(v_i)$ do not interfere
with one another.

\section{Flattening the Halves.}
The next step is to flatten each chosen
half $P_1$ and $P_2$ (independently) 
by suturing in ``curvature triangles''
along each $\sp(v)$ path.
Let $P$ be one of $P_1$ or $P_2$.
The basic idea goes back to Alexandrov~\cite{a-cp-05}[p.~241, Fig.~103],
and was used also in~\cite{iv-cfpcs-08}.
Let $\ell$ be the length $|\sp(v)|$ of a shortest path,
and let $\o=\o(v)> 0 $ be the curvature at $v$.
We glue into $\sp(v)=v v'$ the isosceles \emph{curvature triangle}
$\triangle$ with apex angle $\o$ gluing
to $v$,
and incident sides of length $\ell$ gluing along the cut $v v'$. 
This is illustrated in
Figure~\figref{cube_geodesic_hemis}, where we show the faces
incident to $Q$ in a planar development in~(b) and~(d),
and after gluing in the triangles in~(c) and~(e).
We display this in the plane for convenience of presentation;
the triangle insertion should be viewed as operations on the manifolds
$P_1$ and $P_2$, each independently.

This procedure only works if $\o < \pi$, for $\o$ becomes the apex of the
inserted triangle $\triangle$.
If $\o \ge \pi$, we glue in two triangles of apex angle $\o/2$, both with their
apexes at $v$.\footnote{
   One can view this as having two vertices with half
   the curvature collocated at $v$.
}
Slightly abusing notation, we use $\triangle$ to represent
these two triangles together.  
In fact we must have $\o < 2\pi$ for any vertex $v$
(else there would be no face angle at $v$),
so $\o/2 < \pi$ and this insertion is indeed well defined.

We should remark that an alternative method of handling $\o \ge \pi$
would be to simply not glue in anything to the 
vertex $v$ with $\o(v) \ge \pi$, in which
case we still obtain the lemma below 
leading to the exact same unfolding.

Now, because $\o$ is the curvature (angle deficit) at $v$, gluing in $\triangle$
there flattens $v$ to have total incident angle $2\pi$.
Thus $v$ disappears as a vertex from $P$ (and two new vertices are created along $Q$).

Let $P^\triangle$ be
the new manifold with boundary after insertion of all 
curvature triangles into $P$.
We want to claim that a planar development of $P^\triangle$ does not
overlap.  This is straightforward for a closed quasigeodesic,
but requires some argument for a quasigeodesic loop.
\begin{lemma}
For each half $P$ of $\P$,
$P^\triangle$ is a planar, simple (non-overlapping) polygon.
\lemlab{convex.polygon}
\end{lemma}
\begin{proof}
$P^\triangle$ is clearly a topological disk: $P$ is, and the insertions of $\triangle$'s
maintains it a disk.
At every interior point of $P^\triangle$, the curvature is zero by construction.
So the interior is flat.

Let $\o_Q$ be the total curvature enclosed within $Q$ on $P$,
and $\tau_Q$ the total turn of $Q$, i.e., the turn of $\bP$.
The Gauss-Bonnet Theorem yields $\tau_Q + \o_Q = 2\pi$.
This is precisely the total turn of $\bP^\triangle$, because
that boundary turns $\tau_Q$, plus a total of $\o_Q$ for all the
inserted curvature triangles.
So indeed the boundary of $P^\triangle$ turns just as much as it should
if it is a planar polygon.  It remains to establish
that it is a simple polygon.

There are two cases to consider,
depending on whether $Q$ is a closed quasigeodesic, or a closed
quasigeodesic loop.
\begin{enumerate}
\item $Q$ is a closed quasigeodesic.
In this case we show that the boundary 
$\bP^\triangle$
is convex.
This follows from the orthogonality of $\sp(v)$ guaranteed by Lemma~\lemref{IIV},
as the base angle of the inserted triangle(s) is
$\pi/2 - \o/2$ for $\o < \pi$,
or $\pi/2 - \o/4$ for $\o \ge \pi$
(see Figure~\figref{Q_convex}; $\o=\o(v)$),
so the new angle is smaller than $\pi$ by $\o/2$ or $\o/4$.
\begin{figure}[htbp]
\centering
\includegraphics[width=0.75\linewidth]{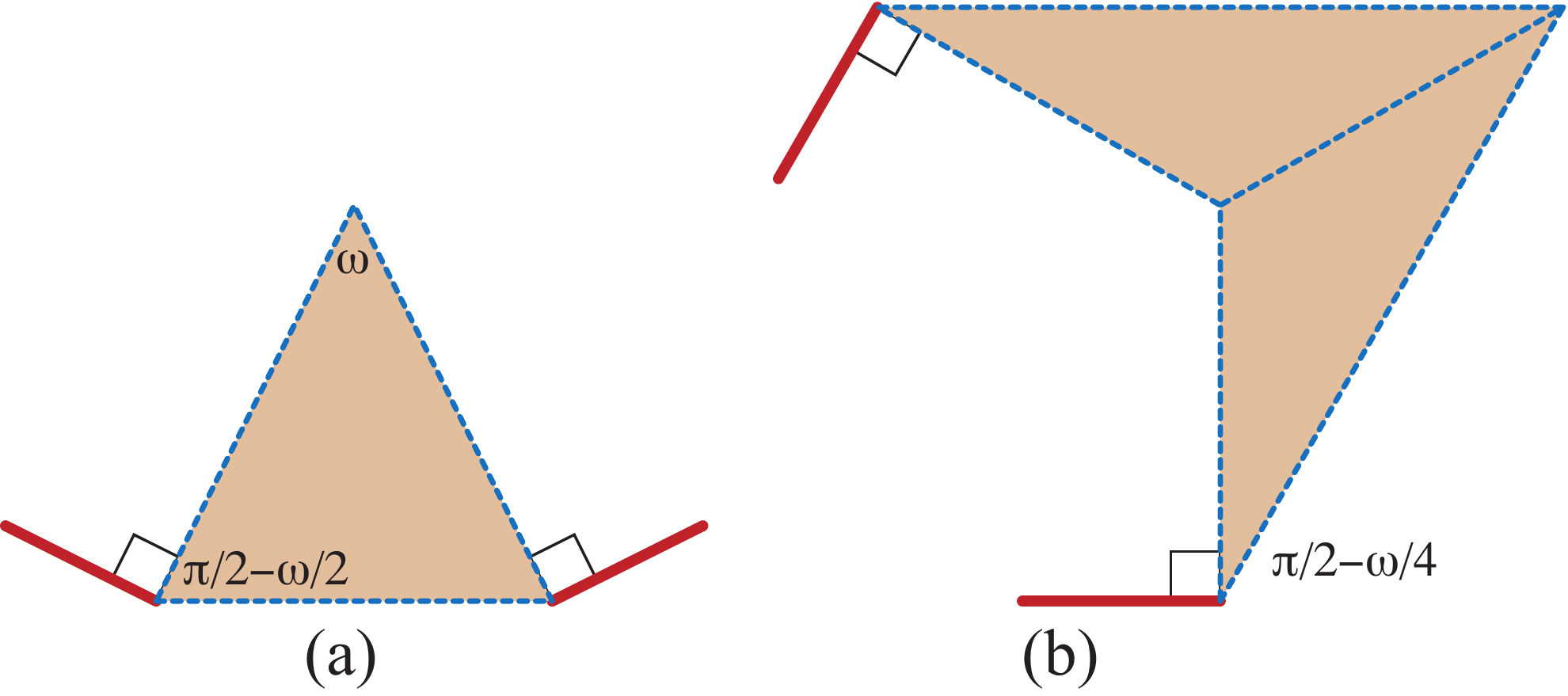}
\caption{Lemma~\protect\lemref{convex.polygon}, Case~1: 
(a)~$\o < \pi$; (b)~$\o \ge \pi$.}
\figlab{Q_convex}
\end{figure}
Thus $P^\triangle$ is a planar convex polygon, and therefore simple.
See Figure~\figref{cube_geodesic_hemis}(c,e) for examples.

\hide{
Note that, when the total curvature in $P$ is $2\pi$
then the straight development
of $Q$ is turned $2\pi$ by the $\triangle$ insertions,
as in~(b) of the figure.
When the total curvature in $P$ is ${<}2\pi$,
the development of $Q$ is not straight, but the $\triangle$ insertions
turn it exactly the additional amount needed to close it to $2\pi$,
as in~(d) of the figure.
}

\item $Q$ is a closed quasigeodesic loop, with loop point $x$.
If $P$ is the half of $\P$ in which the angle at $x$ is ${<}\pi$,
then the argument above applies.
So assume $P$ is the half in which the angle $\b$ at $x$ exceeds $\pi$.
We consider two subcases.

\begin{enumerate}
\item No vertex of $P$ projects to $x$.

Then after insertion of the curvature triangles, $P^\triangle$ is a topological
disk whose boundary is locally convex at all points except at $x$,
whose internal angle is $\b > \pi$.
We now argue that a planar development of such a domain is non-overlapping.
Let $R_1$ and $R_2$ be rays from $x$ along the two edges of $P^\triangle$
incident to $x$;
see Figure~\figref{xNeighborhoodNov}.
\begin{figure}[htbp]
\centering
\includegraphics[width=0.5\linewidth]{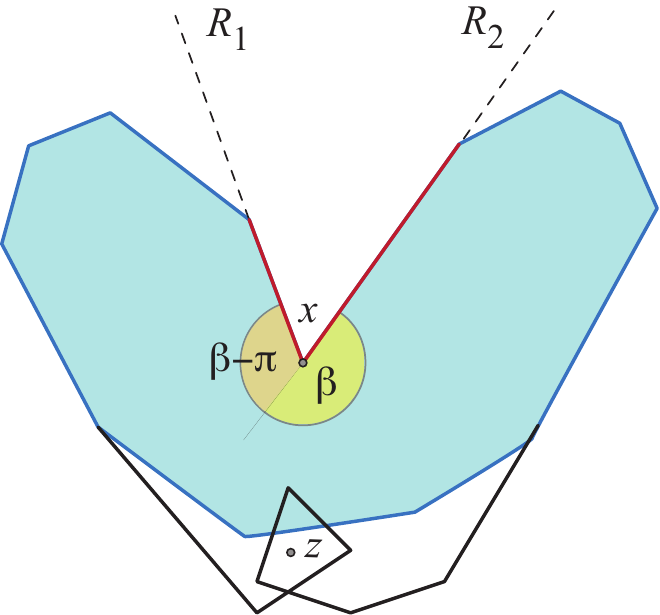}
\caption{Lemma~\protect\lemref{convex.polygon}, Case~2(a): 
$P^\triangle$ has one point $x$ of local nonconvexity.
Here $\b=305^\circ$.
The spiral depicted encloses a point $z$ of winding number 2.
}
\figlab{xNeighborhoodNov}
\end{figure}
The boundary $\bP^\triangle$ of $P^\triangle$ must be exterior to the cone
delimited by $R_1$ and $R_2$ in a neighborhood of those rays,
because the boundary turns convexly at each boundary vertex.
So now we have a convex curve, a subset of $\bP^\triangle$,
leaving $R_1$ and returning to $R_2$.
For the purposes of contradiction, assume this curve self-intersects.
Then it must ``spiral,'' enclosing a point $z$ of winding number ${\ge}2$.
We noted above that the total turn of $\bP^\triangle$ is $2\pi$.
Thus the total turn of
the convex portion of $\bP^\triangle$, i.e.,
$\bP^\triangle \setminus \{x\}$,
exceeds $2\pi$ by the amount $0 <  \b-\pi$ needed
to close the shape at $x$.  
But $\b-\pi < \pi$, so
the convex curve
turns at most $2\pi+ (\b-\pi) < 3\pi$.
However, the point $z$ must ``see'' a turn of ${\ge}4\pi$ to have
winding number ${\ge}2$.
Therefore,  $\bP^\triangle$ does not selt-intersect,
and  $P^\triangle$ is a simple polygon.

\item One or more vertices of $P$ project to $x$.

\hide{
\begin{figure}[htbp]
\centering
\includegraphics[width=\linewidth]{Figures/quasicones}
\caption{Lemma~\protect\lemref{convex.polygon}, Case~2(b): 
(a)~Several vertices project to $x$. 
(b,c,d)~Insertion of curvature triangles. Here
$\o(v_i)=10^\circ$ for $i{=}1,2,3$.
}
\figlab{quasicones}
\end{figure}
}
\begin{figure}[htbp]
\centering
\includegraphics[width=0.75\linewidth]{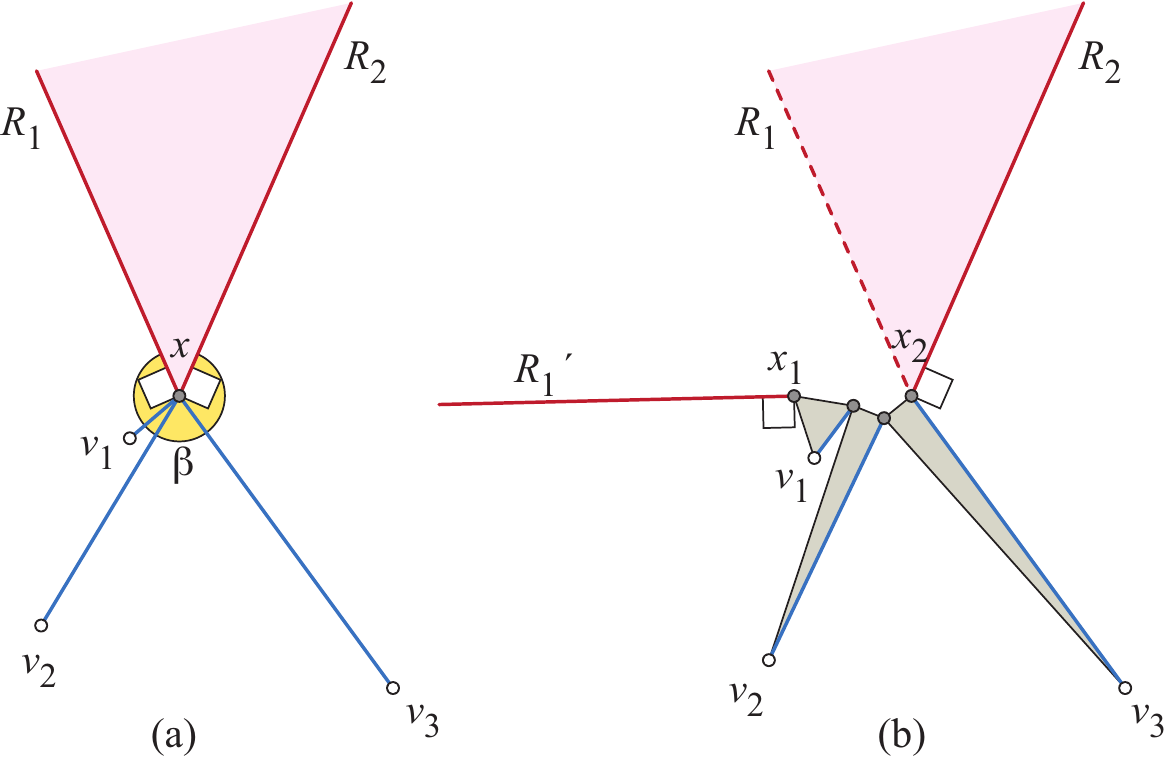}
\caption{Lemma~\protect\lemref{convex.polygon}, Case~2(b): 
(a)~Several vertices project to $x$.
(b)~After insertion of curvature triangles, with $R_2$ held fixed. $R_1'$ is the
planar image of $R_1$.
}
\figlab{NonconvexChain}
\end{figure}

Let $v_1,v_2,\ldots,v_k$ be the vertices that project to $x$
in circular order,
as illustrated in Figure~\figref{NonconvexChain}(a).
\hide{
Insert the curvature triangles one by one, for $i=1,\ldots,k$.
The first rotates the ``elbow''
$R_1xv_1$ by $\o(v_1)$ about $v_1$ to an elbow with ray $A_1$,
the second insertion rotates the second elbow by $\o(v_2)$ about $v_2$,
and so on, as illustrated in~(b,c,d) of the figure.
Let $\a_i$ be the elbow angle at $x$, the angle between $R_1$ and $xv_i$ on $P$.
It must be that $\a_i \ge \pi/2$,
because $v_i x$ is a shortest path.
This angle condition ensures that 
successive elbows do not intersect, and so they 
bound between them ``quasicones'' with disjoint interiors, 
and disjoint
from the original cone delimited by $R_1$ and $R_2$,
which remains after the final rotation~(d).
}
Let $x_1$ and $x_2$ be the extreme images of $x$ in a planar
development of $P^\triangle$
after insertion of all curvature triangles, i.e.,
incident to the planar image of $R_1$ and of $R_2$ respectively;
see Figure~\figref{NonconvexChain}(b).
View the curve $\bP^\triangle$ as composed of two pieces:
$C_{1,2}$, the curve counterclockwise from $x_1$ to $x_2$,
and $C_{2,1}$, the complementary curve counterclockwise from $x_2$
to $x_1$; $\bP^\triangle = C_{1,2} \cup C_{2,1}$.
$C_{1,2}$ is locally convex everywhere, but
$C_{2,1}$ may be nonconvex, as illustrated in Figure~\figref{NonconvexChain}(b).
Now we partition the remaining argument into three parts.

\begin{enumerate}
\item $C_{1,2}$ does not self-intersect.
The convex portion does not self-intersect for the same
reason we just established in the case above:
the curve would have to spiral and turn ${\ge}4\pi$, but
that total turn angle is not available.

\item $C_{2,1}$ does not self-intersect.
It clearly cannot if $k{=}1$, so we
henceforth assume $k \ge 2$.
\begin{figure}[htbp]
\centering
\includegraphics[width=0.5\linewidth]{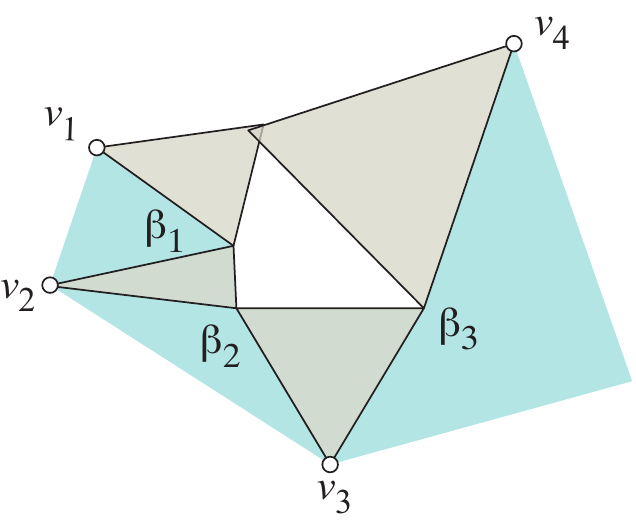}
\caption{Lemma~\protect\lemref{convex.polygon}, Case~2(b)ii: 
The curve $C_{2,1}$ is formed from the bases
of curvature triangles. It needs to turn $\pi$ to self-intersect.}
\figlab{x2x1}
\end{figure}

This portion of $\bP^\triangle$ is composed entirely of bases of
curvature triangles.  For this portion to self-intersect, it must
turn at least $\pi$.  We now compute the total turn $\tau_{2,1}$ and show this
leads to a contradiction.
Let $\a_1$ be the angle $R_1 x_1 v_1$ and $\a_2$ the angle $R_2 x_2 v_k$.
It must be that $\a_i\ge \pi/2$ because $v_ix$ is a shortest path.
Let $\b_i$ be the angle at $x$ on $P$ between $v_i x$ and $v_{i+1} x$.
See Figure~\figref{x2x1}.
Thus,
because $\a_1,\a_2 \ge \pi/2$,
$\a_1 + \a_2 = \b -\sum_{i=1}^k \b_i \geq \pi$, 
and because
$\sum_{i=1}^k \b_i + \a_1 + \a_2 = \b < 2\pi$ 
we get
$\sum_{i=1}^k \b_i < \pi$.
We henceforth drop the limits on the sums, which all run
the full appropriate range.
The base angle for the curvature triangle incident to $v_i$
is $\frac{1}{2}(\pi-\o(v_i))$.
Thus the turn of the curve at the angle $\b_i$ is 
$$
[\frac{1}{2}(\pi- \o(v_i)) + \frac{1}{2}(\pi- \o(v_{i+1})) + \b_i] -\pi
= \b_i - \frac{1}{2}[\o(v_i) + \o(v_{i+1})] .
$$
The total turn is therefore
$$
\tau_{2,1} = \sum \left\{ \b_i - \frac{1}{2}[\o(v_i) + \o(v_{i+1})] \right\}
 = \sum \b_i - \sum \o(v_i).
$$

To insist that $\tau_{2,1} {>} \pi$ is to say that
$$0< \sum \o(v_i) < \sum \b_i - \pi$$
a contradiction to $\sum \b_i - \pi < 0$.
Thus $\tau_{2,1} \le \pi$.
It is also worth noting that the same turn-angle bound holds
for any subchain of $C_{2,1}$ (just by narrowing the sum limits),
a fact we will use below.

\item
$C_{1,2}$ and $C_{2,1}$ do not intersect.
For $C_{2,1}$, a potentially nonconvex curve, to intersect $C_{1,2}$,
it would have to form a path from $x_1$, crossing $R_1'$
(see Figure~\figref{NonconvexChain}(b)), and returning to $x_2$
(or symmetrically, from $x_2$ cross $R_2$ and return to $x_1$).
But this requires some subchain of $C_{2,1}$ to turn ${>}\pi$,
contradicting the turn-angle conclusion above.
Therefore, $\bP^\triangle$ does not self-cross,
and $P^\triangle$ is indeed a simple polygon.
\end{enumerate}

\end{enumerate}
\end{enumerate}
\end{proof}

\noindent
The above argument would be simpler if it were established that
the cone $R_1xR_2$ in  Figure~\figref{NonconvexChain}(b)
is empty in the planar development.
We leave this for future work.

\section{Joining the Halves}
The third and final step of the unfolding procedure
selects a \emph{supporting segment} $s \subset Q$
whose relative interior does not contain
a projection $v'$ of a vertex.
All of $Q$ will be cut except for $s$.
Our choice of $s$ depends on whether $Q$ is a closed quasigeodesic
or a quasigeodesic loop:
\begin{enumerate}
\squeezelist
\item $Q$ is a closed quasigeodesic.
Then any $s$ generates a supporting line to a planar development of $P_i$,
$i=1,2$,
because $P_i^\triangle$ is a convex domain.
Then joining planar developments of $P_1$ and $P_2$ along $s$
places them on opposite sides of the line through $s$,
thus guaranteeing non-overlap.
See Figure~\figref{cube_geodesic_hemis}(f),
where $s=v_5 v'_6$.
\item $Q$ is a quasigeodesic loop.
Let $P$ be the half of $\P$ that contains the angle $\b > \pi$ at $x$.
Thus $P^\triangle$ is potentially nonconvex at
points along the chain $C_{2,1}$ from $x_2$ to $x_1$.
\hide{
Nevertheless, some $s$ might still be a supporting line of
$P_1$.
This is the case in Figure~\figref{GeodesicLoopCube}, where several
supporting $s$ options exist.

Although we do not have an example where no such supporting
$s$ exists, neither do we have a method to guarantee that one
does exist.
So we follow a different strategy.

The argument using disjoint quasicone interiors 
in Lemma~\lemref{convex.polygon}
established that the complementary cone at
$x$
of angle $2\pi-\b$, cone $R_1x_2R_2$ in 
Figure~\figref{quasicones}(d), is empty in the planar
development of $P_1$.
Thus this represents a \emph{supporting cone} to $P_1$.
Now $P_2^\triangle$ is a convex planar domain, whose angle
at $x$ is $2\pi-\b$.  Thus the planar development of
$P_2$ may be placed so that $x$ of $P_2$ and $x_2$ of $P_1$ coincide,
and $s$ is the segment incident to $x_2$.
In Figure~\figref{GeodesicLoopCube}(b),
$s=x_2 v'_2$.
Note that, due to symmetry, the segment $s$ incident to $x_1$
is an equally valid choice.
}
Let $R_1$ and $R_2$ be the rays from $x_1$ and $x_2$ respectively, tangent
to 
$\bP^\triangle$.
\begin{figure}[htbp]
\centering
\includegraphics[width=0.8\linewidth]{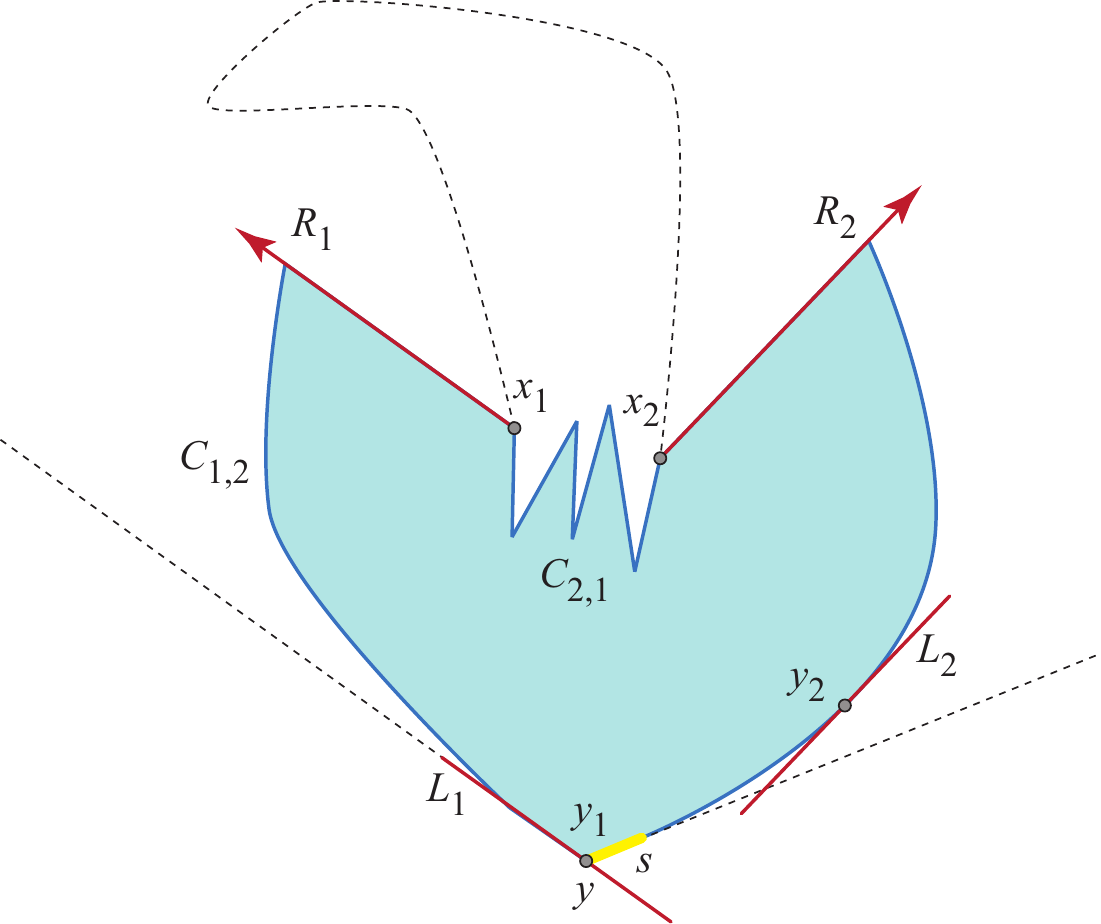}
\caption{
$C_{2,1}$ cannot cross both extensions of 
edges incident to $y$.
Here $y=y_1$.
}
\figlab{SupportingLine}
\end{figure}
Let $L_1$ and $L_2$ be lines parallel to $R_1$ and $R_2$
tangent to $\bP^\triangle$ at $y_1$ and $y_2$ respectively.
Let $y$ be any vertex between $y_1$ and $y_2$;
$y$ may be $y_1$ or $y_2$, as in
Figure~\figref{SupportingLine}.
Now we claim that one of the two edges of 
$\bP^\triangle$
incident to $y$ can serve as $s$.
For both these edges to fail to extend to supporting
lines, $C_{2,1}$ would have to cross both edge extensions.
But, the angle at $y$ is ${<}\pi$
(because $C_{1,2}$ is convex),
so crossing both edge extensions would require
$C_{2,1}$ 
to turn more than $\pi$, which we established in
Lemma~\lemref{convex.polygon} is impossible.

Case~2 is illustrated in Figure~\figref{GeodesicLoopCube}(b),
where this reasoning leads to $s=v'_0 v'_7$.
\end{enumerate}

\noindent
In either case $s$ extends to a supporting line
of both halves, and thus we obtain
a non-overlapping placement of the
planar developments of $P_1$ and $P_2$.

It should be clear now that this procedure works for any convex polyhedron:
\begin{theorem}
Let $Q$ be a quasigeodesic loop on a convex polyhedral surface $\P$.
Cutting shortest paths from every vertex to $Q$, and cutting all but
a supporting segment $s$ of $Q$ as designated above,
unfolds $\P$ to a simple planar polygon.
\thmlab{main}
\end{theorem}

\begin{figure}[htbp]
\centering
\includegraphics[width=0.8\linewidth]{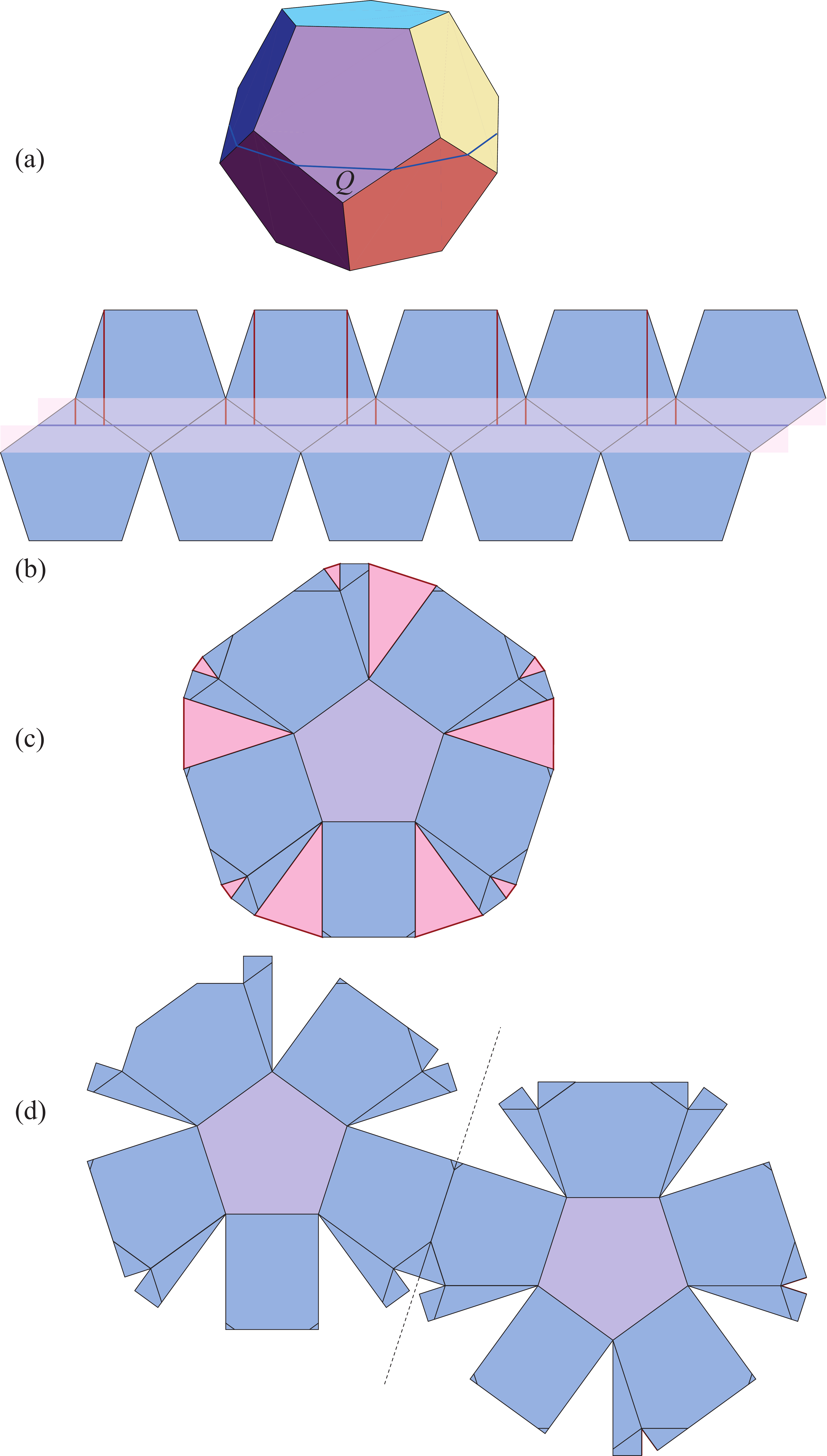}
\caption{(a)~$Q$ here is a geodesic; it includes no vertices, as is
evident in the layout~(b).
The region isometric to a right
circular cylinder is highlighted.
The convex domain $P^\triangle$
from Lemma~\protect\lemref{convex.polygon}
is shown in~(c),
and one possible unfolding in~(d).
}
\figlab{DodecahedronUnfolding}
\end{figure}

Figure~\figref{DodecahedronUnfolding} shows another
example, a closed geodesic on a dodecahedron, this time a pure geodesic.
The unfolding following the above construction is shown
in Figure~\figref{DodecahedronUnfolding}(c,d).
In this case when
$Q$ is a pure, closed geodesic, there is additional structure that
can be used for an alternative unfolding.
For now $Q$
lives on a region isometric to a right
circular cylinder. 
Figure~\figref{DodecahedronUnfolding}(b)
illustrates that
the upper and lower rims of the cylinder are loops
parallel to $Q$ through the vertices of $P$ at minimum distance
to $Q$ (at least one vertex on each side.) 
In the figure, 
these
shortest distances to the upper rim are the short vertical paths from $Q$
to the five pentagon vertices.
Those rim loops are themselves
closed quasigeodesics.
An alternative unfolding keeps the cylinder between the rim loops intact
and attaches the two reduced halves to either side.
See Figure~\figref{DodecaCostin}.
\begin{figure}[htbp]
\centering
\includegraphics[width=0.75\linewidth]{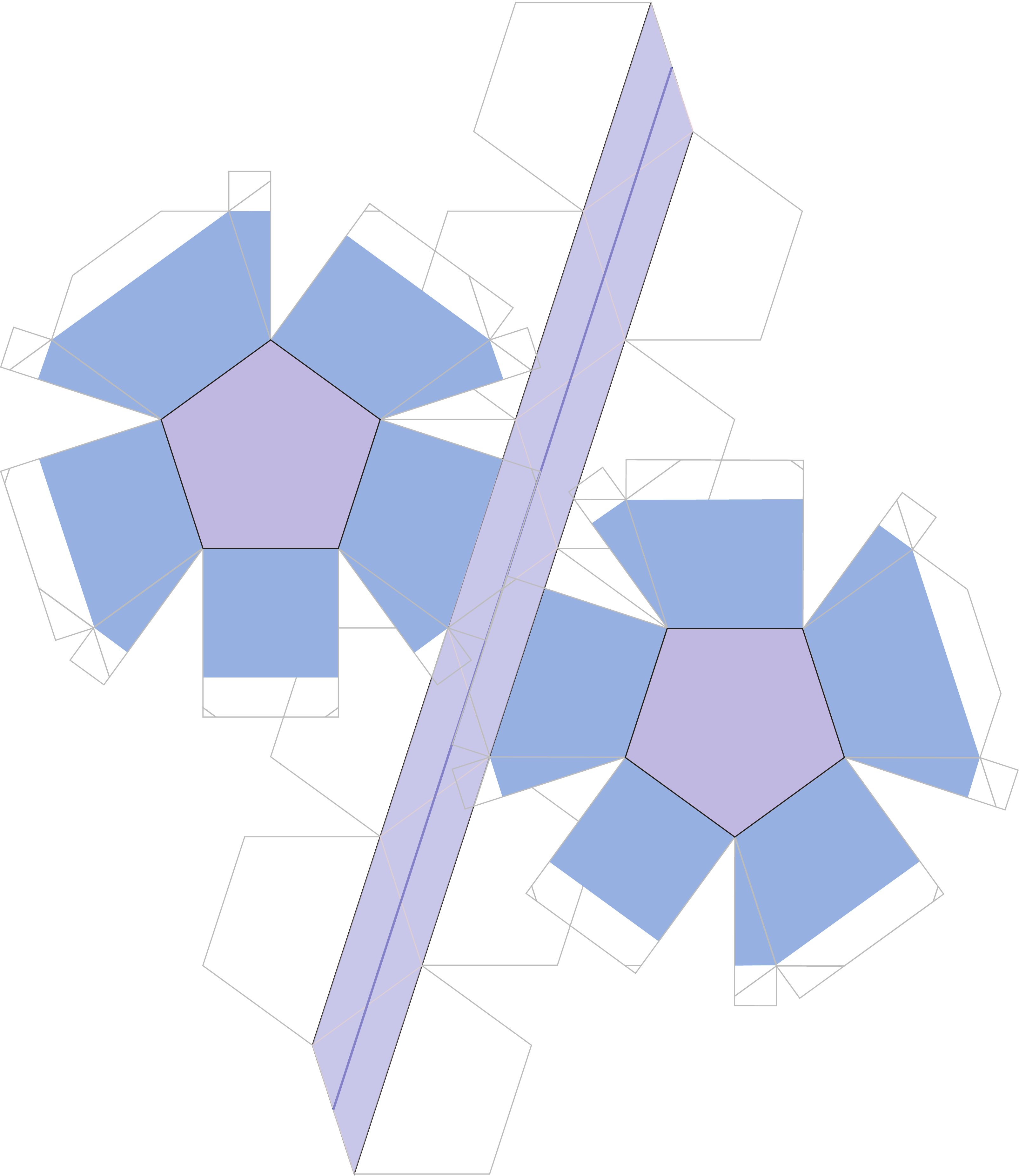}
\caption{Alternative unfolding of the example in
Figure~\protect\figref{DodecahedronUnfolding}.
Various construction lines are shaded lightly.
}
\figlab{DodecaCostin}
\end{figure}

\section{Future Work}
We have focused on establishing Theorem~\thmref{main} rather than the algorithmic
aspects.  Here we sketch preliminary thoughts on computational complexity.
Let $n$ be the number of vertices of $\P$, and
let $q=|Q|$ be the number of faces crossed by the
geodesic loop $Q$.
In general $q$ cannot be bound as a function of $n$.
Finally, let $m=n+q$, the total combinatorial complexity
of the ``input'' to the algorithm.
Constructing $Q$ from a given point and direction will
take $O(q)$ time.
Identifying a supporting segment $s$, and laying out the final
unfolding, is proportional to $m$.
The most interesting algorithmic challenge is to find the shortest
paths from each vertex $v_i$ to $Q$.
It appears that this can be accomplished efficiently,
in $O(m \log m)$ time, by first computing the cut locus of $Q$.
We expect to address this computation in~\cite{iov-ucpqsu-08b}.

We do not believe that quasigeodesic loops constitutes the widest class of
curves for which the star unfolding leads to non-overlap.
In particular, we believe we can extend
Theorem~\thmref{main} to
quasigeodesics with two exceptional points, one with angle ${>}\pi$
to one side, and the other with angle ${>}\pi$ to the other side.
But whether this extension constitutes the widest class of curves for which
Theorem~\thmref{main} holds remains unclear. 

If one fixes a nonvertex point $p \in \P$ and a surface direction $\overrightarrow{u}$ 
at $p$, a quasigeodesic loop 
can be generated to have direction $\overrightarrow{u}$ at $p$.
It might be interesting to study the continuum of star unfoldings 
generated by spinning $\overrightarrow{u}$ around $p$.

\section*{Appendix~1: Symbol Glossary}
\begin{tabular}{l l}
$\P$ 
   & convex polyhedron \\
$P_1, P_2$ 
   & the two ``halves'' $\P \setminus Q$ \\
$P$
   & one half, either $P_1$ or $P_2$ \\
$v_i$ 
   & vertex of $\P$ \\
$Q$ 
   & a quasigeodesic loop \\
$x$
   & the exceptional loop point of $Q$ \\
$L(p),R(p)$
   & angle incident to left/right side of $p$ on curve \\
$\G$
   & directed curve \\
$\sp(v)$
   & one selected shortest path from $v$ to $Q$ \\
$v'$
   & the projection of $v$ onto $Q$: $\sp(v) = v v'$ \\
$\o=\o(v)$
   & the curvature at $v$, $2\pi$ minus the incident face angles \\
$\triangle$
   & a curvature triangle \\
$\ell$
   & $|\sp(v)|$ \\
$\bP$
   & the boundary of $P$ \\
$P^\triangle$
   & the manifold $P$ after insertion of all curvature triangles $\triangle$ \\
$\o_Q$
   & total curvature inside $Q$ on $P$ \\
$\tau_Q$
   & total turn of $Q = \bP$ \\
$\bP^\triangle$
   & the boundary of  $P^\triangle$ \\
$\b$
   & angle ${>}\pi$ at the loop point $x \in Q$ \\
$x_1,x_2$
   & extreme images of $x$ in planar development of $P^\triangle$ \\
$R_1,R_2$
   & rays along edges incident to $x_1,x_2$\\
$\a_i$
   & $\a_1$ is angle $R_1 x_1 v_1$ and $\a_2$ is angle $R_2 x_2 v_k$\\
$\b_i$
   & angle at $x$ on $P$ between $v_i x$ and $v_{i+1} x$ \\
$C_{1,2}$
   & $\bP^\triangle$ counterclockwise from $x_1$ to $x_2$ \\
$C_{2,1}$
   & $\bP^\triangle$ counterclockwise from $x_2$ to $x_1$\\
$\tau_{2,1}$
   & total turn of curve $C_{2,1}$ from $x_2$ to $x_1$\\
$L_1,L_2$
   & lines parallel to $R_1,R_2$ tangent to $P^\triangle$\\
$y_1,y_2$
   & tangency points of $L_1,L_2$\\
$y$
   & a vertex between $y_1$ and $y_2$\\
$s$
   & supporting segment\\
$n$
   & number of vertices of $\P$ \\
$q$
   & combinatorial complexity of $Q$\\
$m$
   & $n+q$\\
$\overrightarrow{u}$ 
   & direction vector through $p \in \P$
\end{tabular}


\bibliographystyle{alpha}
\bibliography{/home/orourke/bib/geom/geom}
\end{document}